\documentclass[conference, 9pt]{IEEEtran}
\usepackage{amsmath,graphicx}
\usepackage{array}
\usepackage{algorithm}
\usepackage{algorithmic}
\usepackage{latexsym}
\usepackage{multirow}
\usepackage{amssymb}
\usepackage[style=base]{caption}
\usepackage{float}
\usepackage{color}
\usepackage{cite}
\usepackage{tabularx}
\usepackage{hyperref}
\usepackage{MnSymbol}
\usepackage{subfig}
\usepackage{graphicx}
\usepackage{epsfig}

\usepackage{slashbox,pict2e}

\renewcommand{\algorithmicrequire}{\textbf{Input:~}}
\renewcommand{\algorithmicensure}{\textbf{Output:~}}

\renewcommand\footnoterule{%
  \kern-0.2cm
  \hrule
  \kern0.2cm}

\IEEEoverridecommandlockouts

\title{Semi-blind Source Separation via Sparse Representations\\
and Online Dictionary Learning}
\author{\IEEEauthorblockN{Sirisha Rambhatla and Jarvis Haupt}
\IEEEauthorblockA{Department of Electrical and Computer Engineering\\
University of Minnesota -- Twin Cities\\
{\tt \{rambh002,jdhaupt\}@umn.edu}. }}
\begin{document}

\maketitle

\begin{abstract}
This work examines a semi-blind single-channel source separation problem.  Our specific aim is to separate one source whose local structure is approximately known, from another \emph{a priori} unspecified background source, given only a single linear combination of the two sources.  We propose a separation technique based on local sparse approximations along the lines of recent efforts in sparse representations and dictionary learning.  A key feature of our procedure is the \emph{online} learning of dictionaries (using only the data itself) to sparsely model the background source, which facilitates its separation from the partially-known source.  Our approach is applicable to source separation problems in various application domains; here, we demonstrate the performance of our proposed approach via simulation on a stylized audio source separation task.
\end{abstract}
\begin{keywords}
Source separation, sparse representations, dictionary learning
\end{keywords}

\section{Introduction}
\label{intro}

The blind source separation (BSS) problem entails separating a collection of signals, each comprised of a superposition of some unknown sources, into their constituent components.  A canonical example of the BSS task arises in the so-called \emph{cocktail party problem} in audio processing, and a number of methods have been proposed to address this problem.  Perhaps the most well-known among these is independent component analysis (ICA) \cite{ICA}, where the sources are assumed to be independent non-Gaussian random vectors. Other source separation approaches entail more classical matrix factorization techniques like principal component analysis (PCA) \cite{
JolliffePCA}, or, when appropriate for the underlying data, non-negative matrix factorization (NMF) \cite{NMF} and non-negative sparse coding (NNSC) \cite{NNSC, NNSC2}.

Here we focus on a slightly different, and often more challenging setting -- the so-called \emph{single channel} source separation problem -- where only a single mixture of the source signals is observed.  Single channel source separation problems require the use of some additional \emph{a priori} knowledge about the sources and their structure in order to perform separation\cite{Jang03, Jung00, Schmidt06, Davies2007}. Here, we assume that the local structure of one of the source signals is \emph{approximately} known (in a manner described in more detail below), and our aim is to separate this partially known source from an unknown ``background'' source.  

Our separation approach is based on local sparse approximations of the mixture data. A novel feature of our proposed method is in our representation of the unknown background source -- we describe a technique for learning (from the data itself) a model that sparsely represents the unknown background source, using tools from the emerging literature on dictionary learning (see, e.g., \cite{Olshausen_Field_1997, KSVD05, DL10}).

While our proposed approach may find utility in source separation tasks in a variety of application domains, our effort here is motivated by an audio processing application in law enforcement scenarios where electroshock devices are utilized to induce temporary incapacitation.  A key forensic task in these scenarios is to determine, from audio data recorded by the device itself, whether the resistive load encountered by the device is in a qualitatively ``high'' or ``low'' state  (corresponding, respectively, to settings where the device is, or is not, delivering current to a human suspect).  We demonstrate our proposed approach on a stylized version of this task, where our approach is employed to separate the audio corresponding to a nominally periodic (up to random timing jitter) and approximately known (up to the resistive load ambiguities) signal from otherwise unknown, but often highly structured, background audio.  In forensics applications the separated audio signal could subsequently be used to classify the state of the resistive load, though here we demonstrate only the separation task, since accurate separation would facilitate accurate classification.

The remainder of this paper is organized as follows.  In Section~\ref{probForm} we state our problem and describe our proposed approach in the context of related existing works, and a detailed description of our proposed method is provided in Section~\ref{algoDiscussion}. We provide an experimental performance evaluation motivated by the aforementioned audio forensics application in Section~\ref{AFapp}, and provide some concluding discussion and remarks in Section~\ref{Conclusions}.


\section{Problem Statement and Related Works}
\label{probForm}

Although the algorithm we develop here may be applied in any of a number of separation tasks, we use the stylized audio source separation example to fix notation and explain our approach.  Let $x\in\mathbb{R}^n$ represent our observed data, and suppose that $x$ may be decomposed as a sum of two sources -- one of which ($x_p\in\mathbb{R}^n$) exhibits local structure that is partially or approximately known, and the other ($x_u\in\mathbb{R}^n$) is unknown. In our motivating audio application for example, $x$ is comprised of samples of an underlying continuous time waveform, and we consider $x_p$ to be samples of a source that is a nominally regular repetition of one of a small number of prototype signals.  Our aim is to separate the sources $x_p$ and $x_u$ from observations of $x$, which may be noisy or otherwise corrupted. 


Our proposed approach is based on the principle of local sparse approximations. In order to state our overall problem in generality, we describe an equivalent model for our data $x$ that facilitates the local analysis inherent to our approach. Let us suppose that $m$ is an integer that divides $n$ evenly, such that $n/m = q$, an integer.  Then $x\in\mathbb{R}^n$ may be represented equivalently as a $m \times q$ matrix $X$:
\begin{equation}\label{eqn:source}
X = X_p + X_u,
\end{equation}
where $X_p$ is a matrix whose columns are non-overlapping length-$m$ segments of $x_p$, and similarly for $X_u$.  The goal of our effort is, in essence, to separate $X$ into its constituent matrices $X_p$ and $X_u$.  

As alluded above, our separation approach entails leveraging local structure in each of the components of $X$. Our main contribution comes in the form of a procedure that, given our ``partial'' information about the columns of $X_p$, enables us to learn in an online fashion and from the data itself a dictionary $D$ such that columns of $X_u$ are accurately expressed as linear combinations of (a small number of) columns of $D$. In a broader sense, our work is related to some classical approximation approaches as well as several recent works on matrix decomposition.  We briefly describe these background and related efforts in matrix decomposition here, in an effort to put our main contribution in context. 

\subsection{Related Works }
\subsubsection{Low Rank and Robust Low Rank Approximation}

Consider the model \eqref{eqn:source} and suppose that the columns of $X_p$ can each be represented as a linear combination of some $r$ linearly independent vectors, implying that $X_p$ is a matrix of rank $r$.  Now, different separation techniques may be employed depending on our assumptions of $X_u$.  Perhaps the simplest case is where the elements of $X_u$ are iid zero-mean Gaussian random variables; in this case, the problem amounts to a denoising problem, which can be solved using ideas from low-rank matrix approximation.  In particular, it is well-known that the approximation $\widehat{X}_p$ obtained via the truncated (to rank $r$) singular value decomposition (SVD) of $X$ is a solution of the optimization 
\begin{equation}\label{SVD}
\widehat{X}_p  =  \underset{L, \ \mbox{rk}(L)\leq r}{\text{arg~min}} \|X-L\|_F^2,
\end{equation}
where $\mbox{rk}(L)$ is the function that returns the rank of $L$, and the notation $\|\cdot\|_F^2$ denotes the squared Frobenius norm, which is the sum of squares of the elements of the matrix.  


It is well-known that certain (non-Gaussian) forms of interference $X_u$ may cause the accuracy of estimators of the low-rank component obtained via truncated SVD to degrade significantly.  This is the case, for example, when $X_u$ is comprised of sparse large (in amplitude) impulsive noise, or contains a few columns that may be construed as outliers in the low-rank model for $X_p$.  Numerous extensions of traditional PCA to these settings have been proposed in the literature; we mention here several recent efforts in robust PCA \cite{RPCA, ChandrasekaranSPW11} which model $X_u$ as a sparse matrix, and aim to simultaneously estimate both the low-rank $X_p$ and the sparse $X_u$, by solving the convex optimization
\begin{equation}
\label{RPCA}
\{\widehat{X}_p, \widehat{X}_u\}  =  \underset{L,S}{\text{arg min}} ~\|L\|_{*} +\lambda \|S\|_{1} \ \ \text{subject to} \ \ X=L+S,
\end{equation}
where $\lambda>0$ is a regularization parameter.  Here $\|L\|_*$ denotes the \emph{nuclear norm} of $L$, which is the sum of the singular values of $L$.  The nuclear norm is a convex relaxation of the non-convex rank function $\mbox{rk}(L)$.  The notation $\|S\|_1$ here denotes the sum of the absolute entries of $S$ -- essentially the $\ell_1$ norm of a vectorized version of $S$, which is a convex relaxation of the non-convex $\ell_0$ quasinorm that counts the number of nonzeros of $S$. 


\subsubsection{Low Rank Plus Sparse in a Known Dictionary}
A useful extension of the robust PCA approach arises in the case where $X_u$ is not itself sparse, but possesses a sparse representation in some known dictionary or basis. One example is the case where the background source is locally smooth, implying it can be sparsely represented using a few low-frequency discrete cosine transform or Fourier basis elements.  Formally, suppose that for some known matrix $D$, we have that $X_u = D A_u$, where the columns of $A_u$ are sparse. The components of $X$ can be estimated by solving the following optimization \cite{LRCS12}
\begin{equation}
\label{LRplusCS}
\{\widehat{X}_p, \widehat{A}_u\}  =   \underset{L,A}{\text{arg min}} ~\|L\|_{*} +\lambda\|A\|_{1} \ \text{subject to} \ X=L+DA
\end{equation}
Note that an estimate $\widehat{X}_u$ of $X_u$ may be obtained directly as $\widehat{X}_u = D \widehat{A}_u$.  This approach assumes (implicitly) a priori knowledge of a dictionary that sparsely represents the background signal, which may be a restrictive assumption in practice.

\subsubsection{Morphological Component Analysis}
A more general model arises when $X_p$ is not low-rank, but instead, its columns are also sparsely represented in a known dictionary.  Suppose that $X_p$ and $X_u$ are sparsely represented in some known dictionaries $D_1$ and $D_2$, such that $X_p = D_1 A_1$ and $X_u = D_2 A_2$, and that the columns of $A_1$ and $A_2$ are sparse.   Such models were employed in recent work on Morphological Component Analysis (MCA) \cite{StarckED05, MicroLoc10, BobinSFMD07}, which aimed to separate a signal into its component sources based on structural differences codified in the columns of the known dictionaries.  The MCA decomposition can be accomplished by solving the following optimization 
\begin{eqnarray}
\label{MCA}
\{\widehat{A}_1, \widehat{A}_2\} &=\underset{A_{1}, A_{2}}{\text{arg min}} ~\|X - D_{1}A_{1} -D_{2}A_{2}\|_{F}^2 \notag\\&~\text{subject to} \ \|A_{1}\|_{1} +\|A_{2}\|_{1}\leq \lambda,
\end{eqnarray}
for some $\lambda > 0$, where the estimates of $X_p$ and $X_u$ are formed as $\widehat{X}_p = D_1 \widehat{A}_1$ and $\widehat{X}_u = D_2 \widehat{A}_2$, respectively. When $X_p$ and $X_u$ are each comprised of a single column, this optimization is equivalent to the so-called Basis Pursuit (or more specifically, Basis Pursuit Denoising) technique \cite{DonohoChen}, which formed a foundation of much of the recent work in sparse approximation.  Note that this approach also assumes a priori knowledge of a dictionary that sparsely represents the background. 

\subsection{Our Contribution:\\
``Semi-blind'' Morphological Component Analysis}

Our focus here is similar to the MCA approach above, but we assume only one of the dictionaries, say $D_1$, is known.  In this case, the MCA approach  transforms into a semi-blind separation problem where we try to also learn a dictionary $D_2$ to represent the unknown signal. Our main contribution comes in the form of a ``Semi-Blind'' MCA procedure, designed to solve the following modified form of the MCA decomposition
\begin{eqnarray}
\label{SBMCAeq}
\{ \widehat{A}_1, \widehat{A}_2,\widehat{D}_2 \} &= \underset{A_{1}, A_{2}, D_{2}}{\text{arg min}}~ \|X - D_{1}A_{1}-D_{2}A_{2}\|_{F}^2 \ \notag \\&~\text{subject to} \ \ \|A_{1}\|_{1}+\|A_{2}\|_{1}\leq \lambda,
\end{eqnarray}
where the columns of the learned $D_2$ are constrained in some way (e.g., so that the $\ell_2$ norms of all columns are bounded by $1$). This modeling approach forms the basis of the remainder of this paper.  


\section{Semi-blind MCA}
\label{algoDiscussion}
As described above, our model assumes that the data matrix $X$ can be expressed as the superposition of two component matrices, $X_p$ and $X_u$.  Further, we assume that each of the component matrices possesses a sparse representation in some dictionary, such that $X_p \approx D_1 A_1$ and $X_u \approx D_2 A_2$, where $D_1$ is known a priori. Our essential aim, then, is to identify an estimate $\widehat{A}_1$ of the coefficient matrix $A_1$ and estimates $\widehat{D}_2$ and $\widehat{A}_2$ of the matrices $D_2$ and $A_2$.  Our estimates of the separated components are then given by $\widehat{X}_p = D_1 \widehat{A}_1$, and $\widehat{X}_u = \widehat{D}_2 \widehat{A}_2$.

We propose an approach to solve \eqref{SBMCAeq} that is based on alternating minimization, summarized here as Algorithm \ref{SBMCA}.  Let $\lambda_1, \lambda_2, \lambda_3 > 0$ be user specified regularization parameters. Our initial estimate of coefficients $A_1$, corresponding to the coefficients of $X_p$ in the known dictionary $D_1$, is obtained via a LASSO-type approach
\begin{equation}
\label{SparseApxEq}
\widetilde{A}_1=\underset{A_1}{\text{arg~min~}}\|X-D_1A_1\|_F^2  + \lambda_1 \|A_1\|_1,
\end{equation}
or other comparable sparse modeling approach, such as orthogonal matching pursuit (OMP) \cite{OMP}.   We then proceed in an iterative fashion, as outlined in the following subsections, for a few iterations or until some appropriate convergence criteria is satisfied. It should be noted that the lack of joint convexity makes it difficult to make global optimality claims for our proposed approach.  In this sense, the overall performance may vary depending on the particular initialization strategy used.

\begin{algorithm}[t]
\caption{Semi-Blind MCA Algorithm}
\label{SBMCA}

\algorithmicrequire {Original Data $X \in \mathbb{R}^{m\times q}$, Known Dictionary $D_1\in\mathbb{R}^{m\times d}$,

\hspace{3em}Regularization parameters $\lambda_1, \lambda_2, \lambda_3 >0$, 

\hspace{3em}Number of elements in unknown dictionary $\ell$.} \vspace{.3em}\\
\textbf{Initialize:}  Obtain $\widetilde{A}_1$ using any sparse approximation strategy\vspace{.6em}\\
\textbf{Iterate (repeat until convergence):}\\
\begin{algorithmic} 
\vspace{-1em}
\REPEAT 
{\STATE \emph{Dictionary Learning:~\\}{~~~~~~$\{\widetilde{D}_2,\widetilde{A}_2\}$  $\leftarrow$ ~} $\underset{D_2,A_2}{\text{arg min}} ~\|X-D_1\widetilde{A}_1-D_2A_2\|_F^2+\lambda_2\|A_2\|_1$  
\STATE \emph{Coefficient Update:~\\} 
~~~~~~~~~~~~~~~~~~~$\widetilde{D} = [D_1 \ \widetilde{D}_2]$\\ 
~~~~~~$[\widetilde{A}_1^T \ \widetilde{A}_2^T]^T \triangleq \widetilde{A} \leftarrow \underset{A}{\text{arg~min}}~\|X-\widetilde{D}A\|_F^2+\lambda_3 \|A\|_1$
}\UNTIL{convergence}\\
\end{algorithmic}

\algorithmicensure {Learned dictionary $\widehat{D}_2 \leftarrow \widetilde{D}_2$,

\hspace{3.6em}Coefficient estimates $\widehat{A}_1 = \widetilde{A}_1, \ \widehat{A}_2 = \widetilde{A}_2.$}
\vspace{0.6em}
\end{algorithm}
\vspace{-0.05cm}
\subsection{Dictionary learning stage}
\label{DLupdate}
Given the estimate $\widetilde{A}_1$, we can essentially ``subtract" the current estimate of $X_p$ from $X$, and apply a dictionary learning step to identify estimates of the unknown dictionary $D_2$ and the corresponding coefficients $A_2$. In other words, we solve
\begin{equation}
\{\widetilde{D}_2, \widetilde{A}_2\} =\underset{D_2,A_2}{\text{arg~min}} ~\|X-D_1\widetilde{A}_1-D_2A_2\|_F^2+\lambda_2\|A_2\|_1.
\end{equation}
Now, given the estimate $\widetilde{D}_2$, we update our current estimate of the \emph{overall} dictionary $\widetilde{D} = [D_1 \ \widetilde{D}_2]$. We then update the \emph{overall} coefficient matrix by solving another sparse approximation problem, as described next.
\vspace{-0.05cm}

\subsection{Sparse approximation stage}
\label{sparseApxSec}
Given our current estimate of the overall dictionary, we update the corresponding coefficient matrices by solving the following LASSO-like problem:
\begin{equation}
\label{SparseApxEq2}
[\widetilde{A}_1^T \ \widetilde{A}_2^T]^T \triangleq \widetilde{A}=\underset{A}{\text{arg~min~}}\|X-\widetilde{D}A\|_F^2 + \lambda_3 \|A\|_1. 
\end{equation}
Now, we extract the submatrix $\widetilde{A}_1$ from $\widetilde{A}$, and repeat the overall processing (beginning with the dictionary learning step).  These steps are iterated until some appropriate convergence criteria is satisfied.

\section{Experimental Evaluation: An Audio Forensics Source Separation Example} 
\label{AFapp}
We demonstrate the performance of our approach on a stylized version of the audio separation task described in the introduction, which is motivated by forensic examination of audio obtained during law enforcement events where electroshock devices are utilized. For the sake of this example, we suppose that the electroshock devices discharge approximately $36$ times per second (a nominal period of $27.8$ ms), and the waveforms generated by the device during discharge take one of two different forms depending on the level of resistive load encountered by the device. 
The collected audio corresponds to the nominally periodic discharge of the device, superimposed with background noise.  Our aim is to separate this superposition into its components. 

\begin{figure}[t]
\centering\epsfig{file=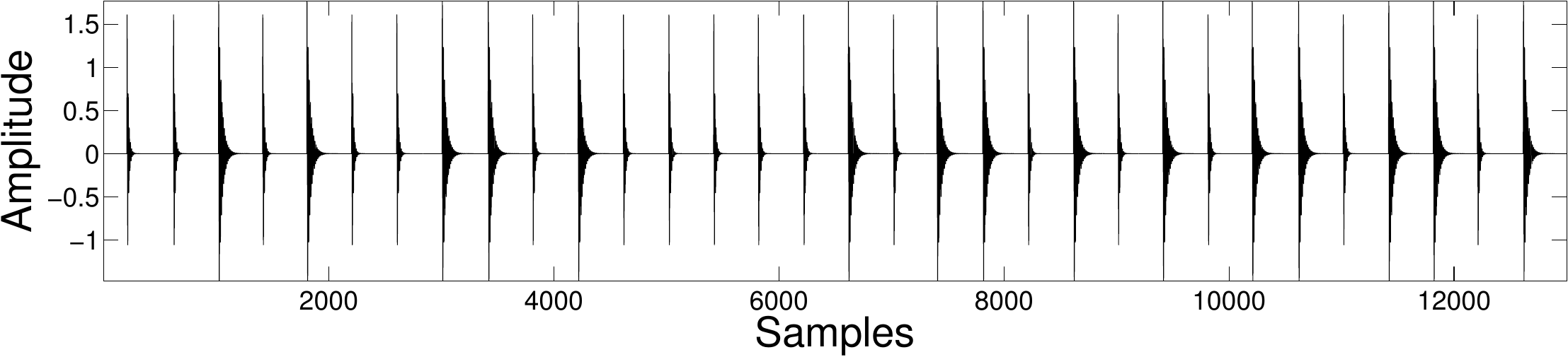,,width=0.97\linewidth,clip=} \\
~~~~~~~(a)\\
\centering\epsfig{file=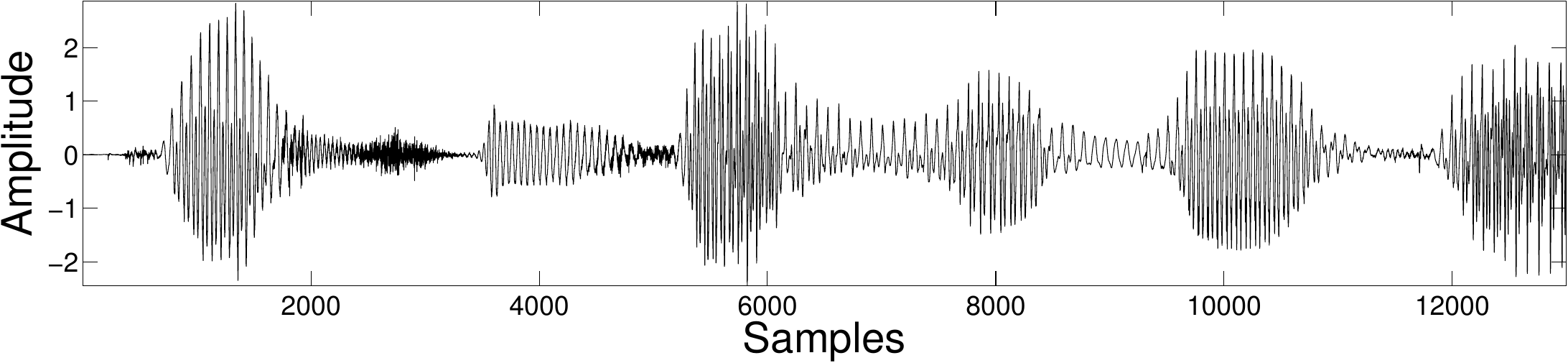,,width=0.97\linewidth,clip=} \\
~~~~~~~(b)\\
\centering\epsfig{file=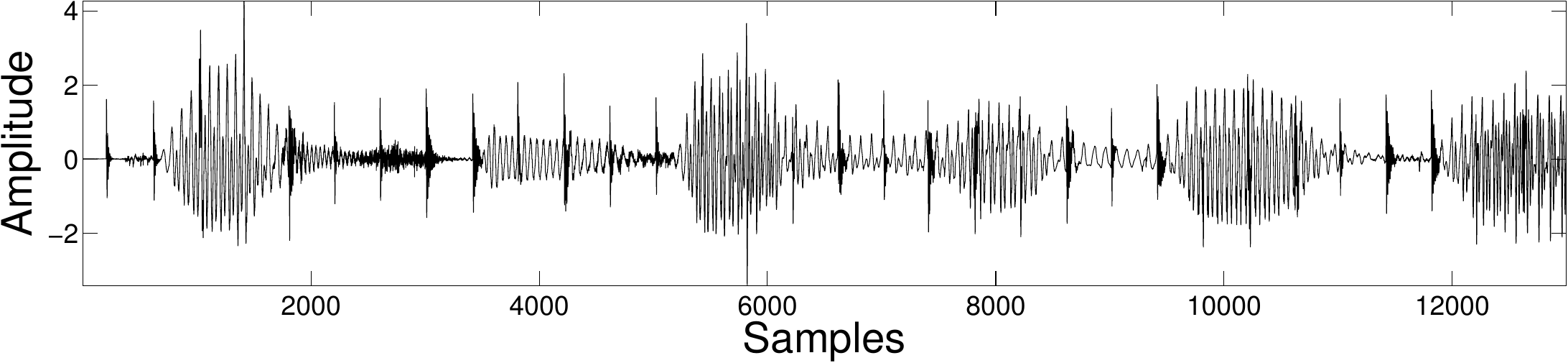,,width=0.97\linewidth,clip=} \\
~~~~~~~(c)
\caption{A segment of mixture components (noise free) used for the experimental validation: (a) the nominally periodic signal $x_p$ (each segment is the discharge corresponding to one of the two resistive load states, randomly selected); (b) the background signal $x_u$; (c) the mixture $x$.}
 \label{Signals}
 
\end{figure}

Figure~\ref{Signals} shows a segment of the signals used in the simulation. We simulate the form of the nominally periodic signals ($x_p$), shown in Figure~\ref{Signals} (a), using two distinct exponentially decaying sinusoids, corresponding to the use of two series RLC circuits with different resistance parameters to model the loaded and open circuit states.  We form the overall signal $x_p$ by concatenating a sequence of randomly-selected versions of these two prototype signals, each of which is subject to a few samples of random timing offset or \emph{jitter}, in order to model the non-idealities of an actual electroshock device\footnote{The timing offsets for each prototype signal were selected randomly from a collection of $20$ distinct values from a symmetric interval $1.39$ ms in duration centered at the nominal pulse occurrence time.}.  A speech signal\footnote{Speech Samples obtained from VoxForge Speech Corpus: \texttt{www.voxforge.org/home}}  shown in  Figure~\ref{Signals} (b), was used to model background source.  We simulate the overall raw audio data as a linear combination of $x_p$,  $x_u$ and zero-mean random Gaussian noise $~\mathcal{N}(0,\sigma^2)$ (Figure~\ref{Signals} (c)  depicts the ideal case $\sigma=0$). The data matrix $X$ is then formed from the signal $x$ as discussed in Section \ref{probForm}, using non-overlapping segments with $400$ samples each. 

Now, we form the dictionary $D_1$ by incorporating $40$ shifts of the nominal prototype pulses, corresponding to $20$ distinct shifts of each pulse, and employ the semi-blind MCA approach (discussed in Section~\ref{algoDiscussion}) to separate the background audio from the approximately known periodic portion.  To evaluate whether, and to what extent, our proposed dictionary learning step aids in the separation relative to a similar approach that uses a fixed dictionary/basis to represent the unknown background source, we compare our proposed approach with two versions of MCA which use fixed bases (the standard DCT basis or the identity basis) to form the dictionary $D_2$.  We refer to these techniques as ``MCA-DCT'' and ``MCA-Identity,'' respectively\footnote{We use the estimated $x_p$, obtained via MCA-DCT procedure to initialize our approach, as follows: we apply one step of orthogonal matching pursuit (OMP)~\cite{OMP} on the estimate of $x_p$ obtained via MCA-DCT to form the initial (one component per column) estimate $A_1$ for the SBMCA algorithm.}.  Further, for this example we evaluate our approach relative to a ``time-frequency masking'' approach, which is a standard method in the audio source separation literature. Specifically, we compare our approach with an approach based on spectral separation using NNSC and NMF, and denote this approach as NNSC (Spectral)\footnote{Let $\check{X}$ denote the matrix whose columns are the non-negative frequency-domain amplitude spectra of the corresponding columns of $X$. Owing to symmetry, we retain in $\check{X}$ only the amplitudes corresponding to positive frequencies. Now, a classical NNSC-based source separation approach identifies $\check{D}$ and $\check{A}$ with nonnegative elements, to minimize $\|\check{X}-\check{D}\check{A}\|_F^2 + \lambda \|A\|_1$ (when $\lambda=0$ this is just NMF). We reconstruct the time-domain estimate of $x_p$ using the spectral information from the two columns of $\check{D}$ having largest total contribution across rows of $\check{A}$, and corresponding phase information from the original mixture (the estimate of $x_u$ is obtained similarly, using the remaining columns of $D$).}.

Table~\ref{SNRper} lists the \emph{best achievable} reconstruction SNRs (in dB)\footnote{For a signal $x_{\rm orig}$ and its estimate $x_{\rm est}$, the SNR is computed as $\mbox{SNR}=20 \log_{10}(\|x_{\rm orig}\|_2/\|x_{\rm est}-x_{\rm orig}\|_2)$.}  for MCA-DCT, MCA-Identity and NNSC (Spectral), obtained by choosing the parameters which give the best performance for $x_p$ and $x_u$ separately. For aforementioned methods, different parameters may have been utilized to obtain the reconstruction SNRs of each signal component, even for the same method and same noise level -- in other words, the SNRs listed may not be jointly achievable from a \emph{single} implementation of these procedures. However in case of SBMCA, the reported SNR values are achieved by \emph{clairvoyantly} tuning the value(s) of the regularization parameters and number of dictionary elements to give the lowest error via a \textit{single} implementation. At any rate, the proposed approach significantly outperforms each of the other approaches (MCA-based, as well as the classical  spectral domain separation) in each of the settings examined. 

\begin{figure*}[t]
\centering
\begin{tabularx}{\textwidth}{@{}lcccc@{}}
~&\multicolumn{2}{c}{~~~~~$\sigma=0$} & \multicolumn{2}{c}{~~~~~~~$\sigma=0.1$} \\
&~~~~~ $x_p$ & ~~~~~$x_u$&~~~~~$x_p$ & ~~~~~$x_u$\\
\raisebox{6.5\height}{SBMCA}&
\epsfig{file=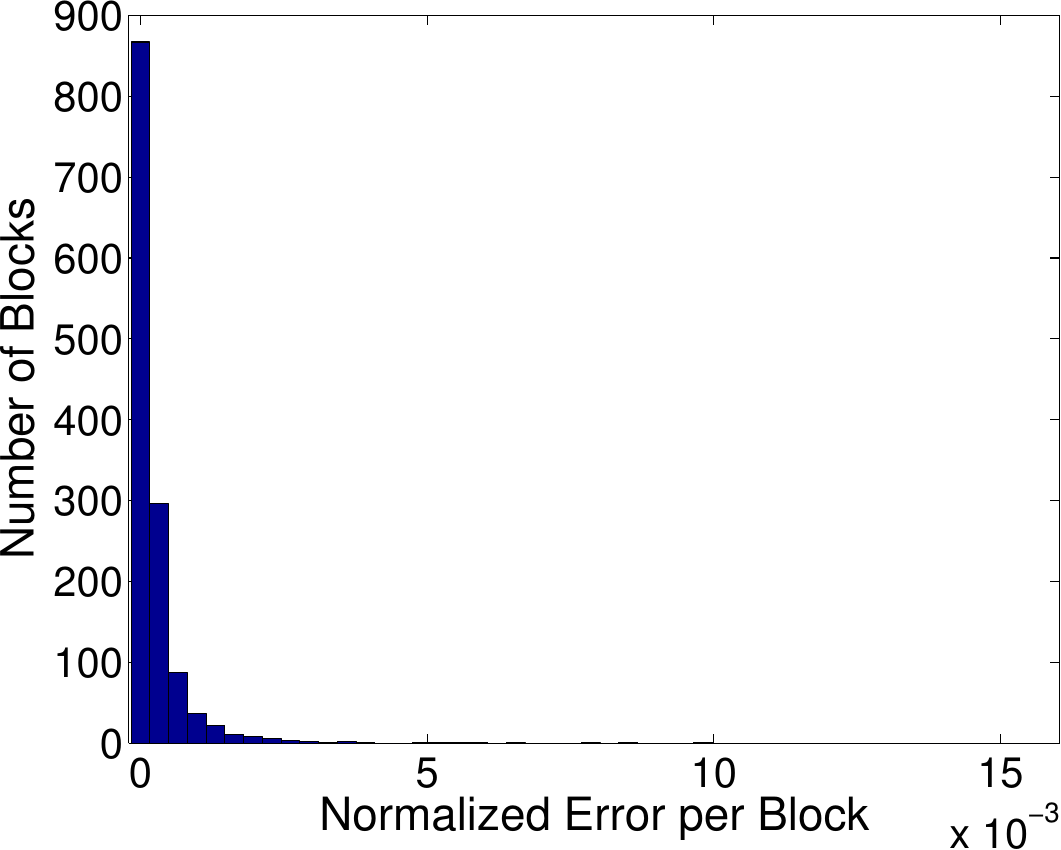,,width=0.19\linewidth,clip=} &
\epsfig{file=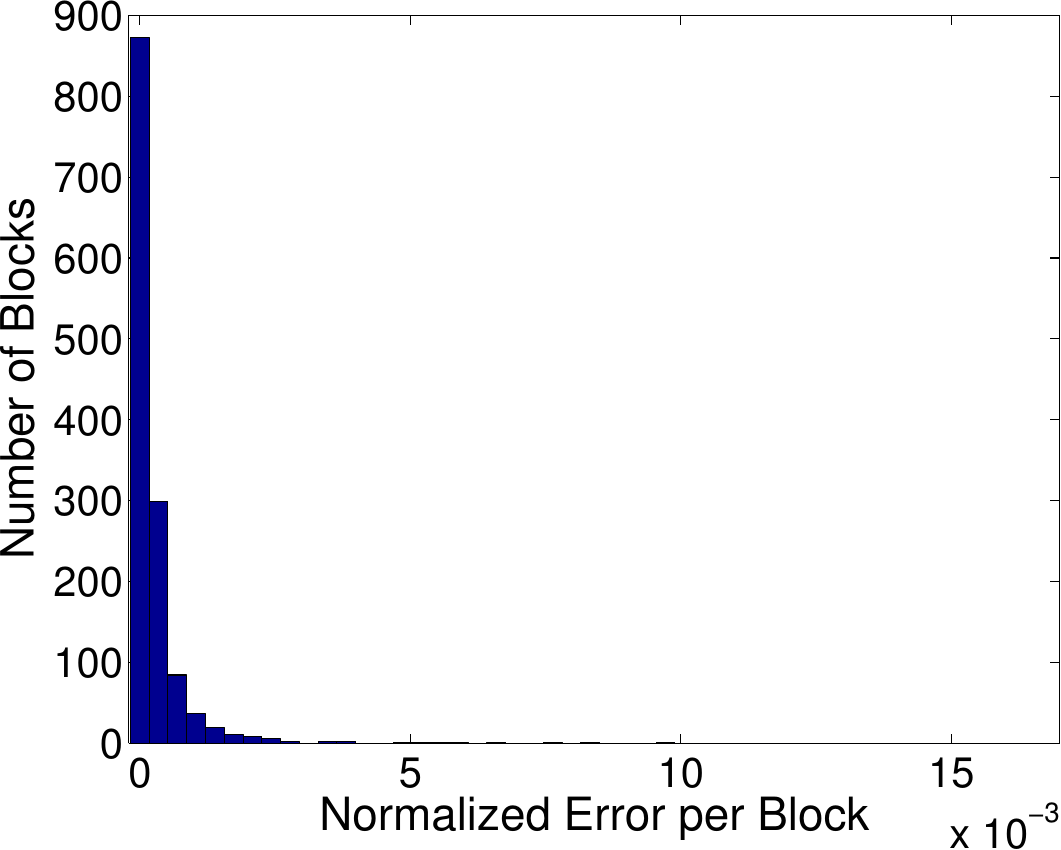,,width=0.19\linewidth,clip=} &
\epsfig{file=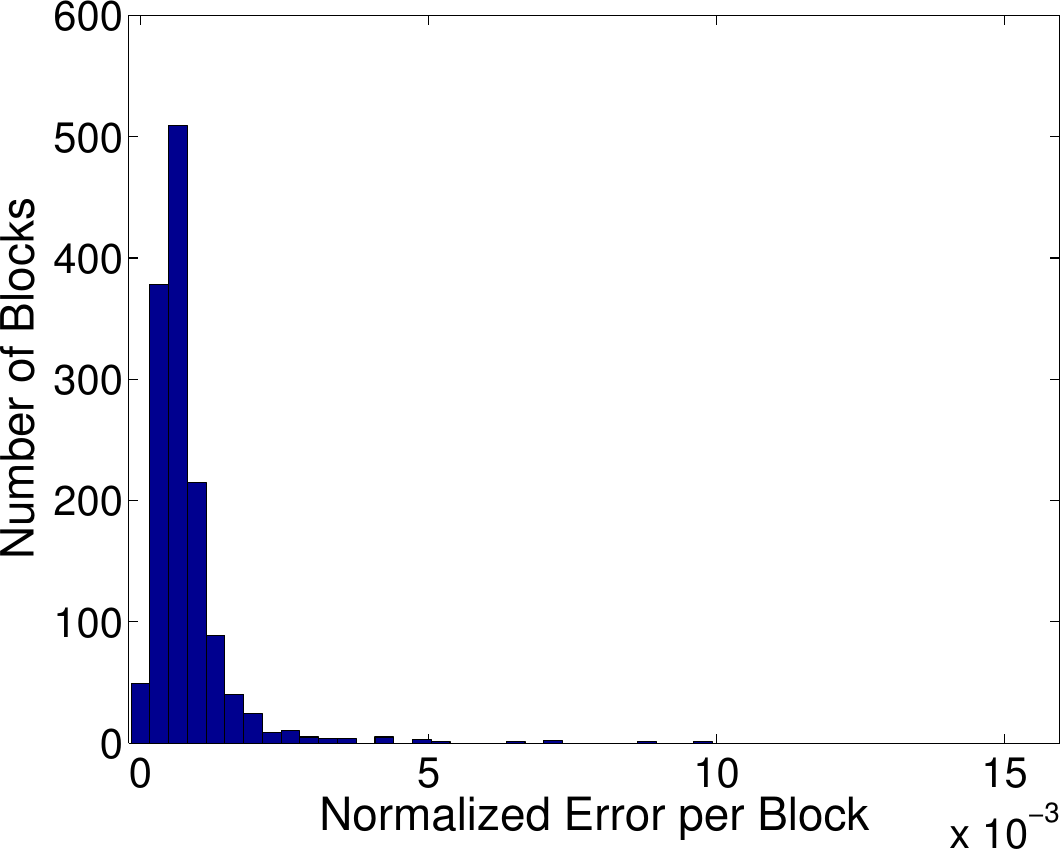,,width=0.19\linewidth,clip=} &
\epsfig{file=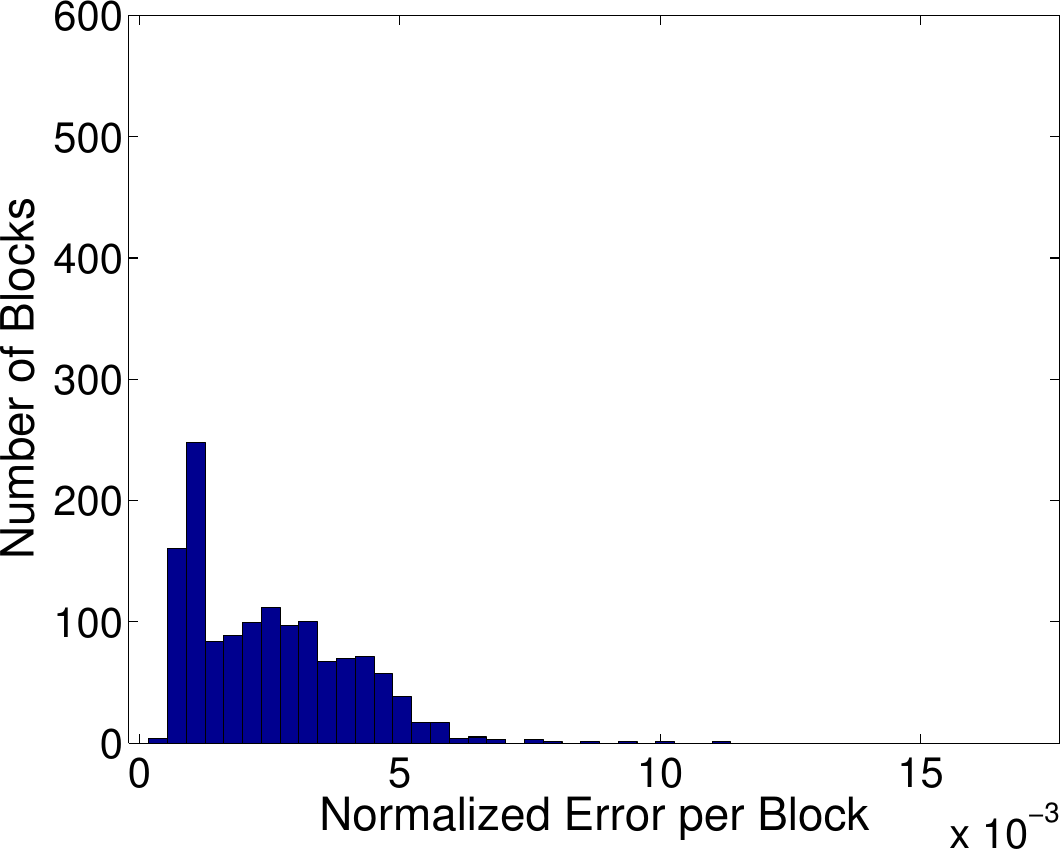,,width=0.19\linewidth,clip=} \\
~&~~~~~~~(a) & ~~~~~~~(b) & ~~~~~~~(c) &~~~~~~~(d)\\
\raisebox{6.5\height}{MCA-DCT}&
\epsfig{file=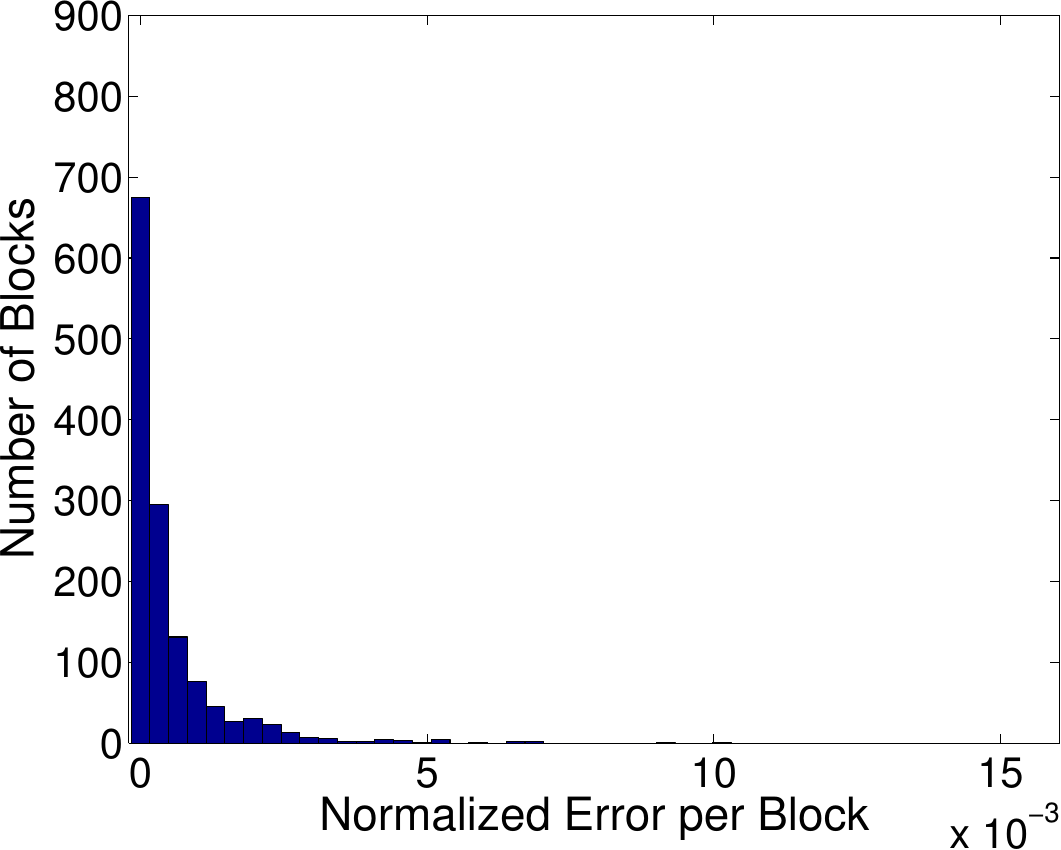,,width=0.19\linewidth,clip=} &
\epsfig{file=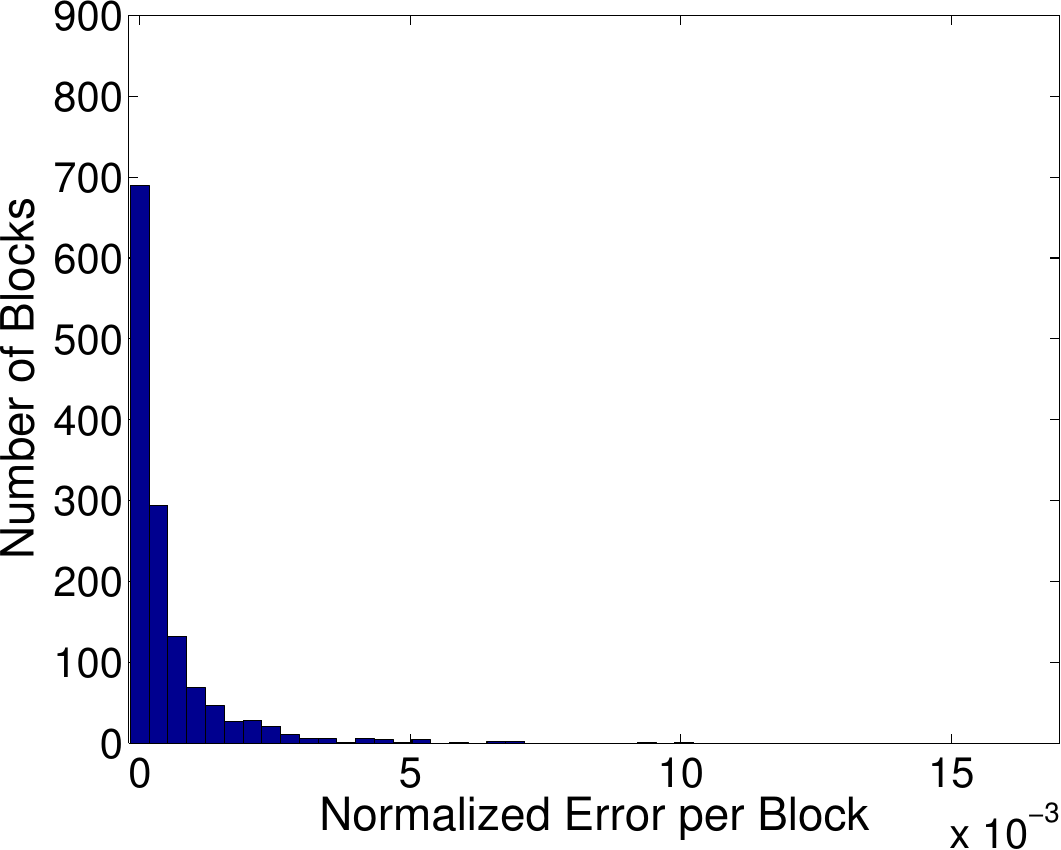,,width=0.19\linewidth,clip=} &
\epsfig{file=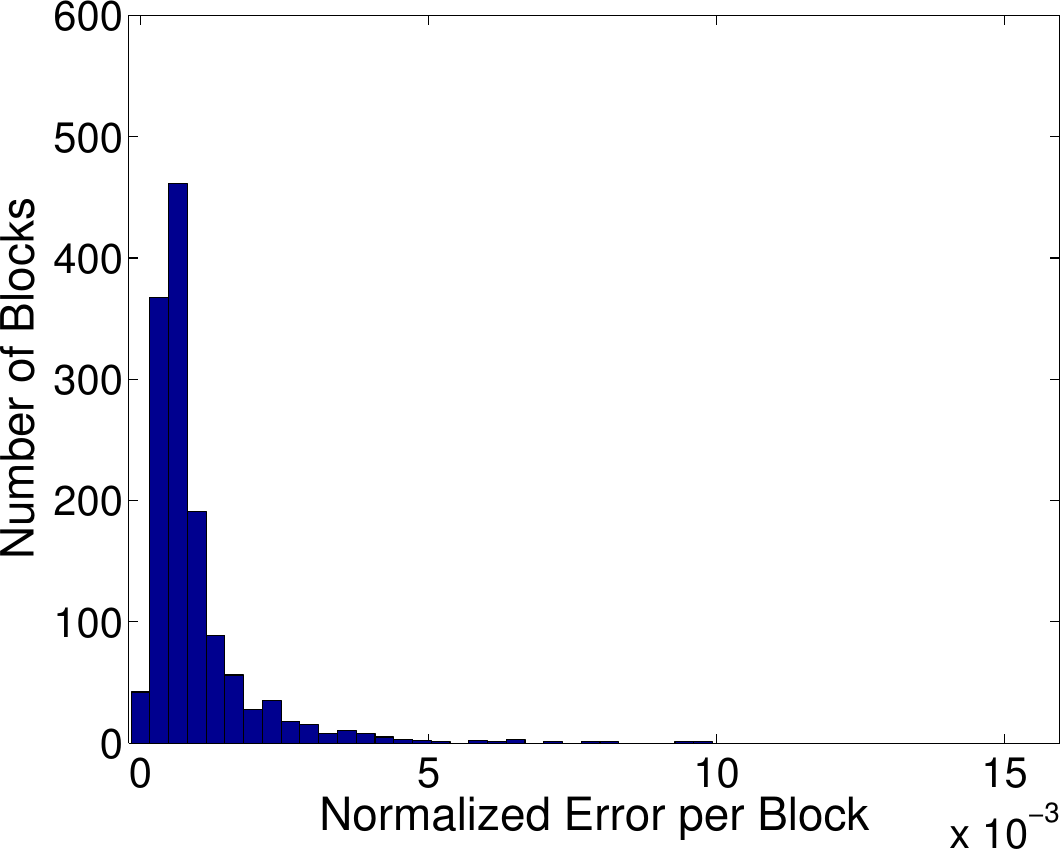,,width=0.19\linewidth,clip=} &
\epsfig{file=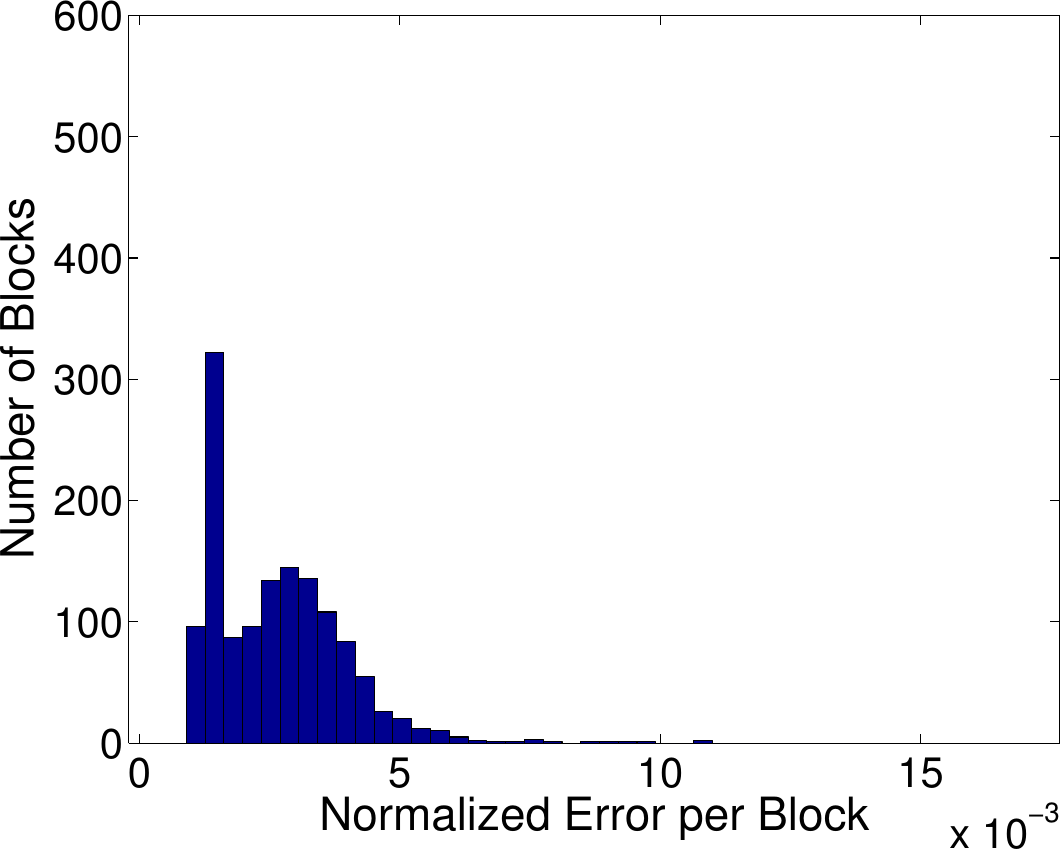,,width=0.19\linewidth,clip=}  \\
~&~~~~~~~(e) & ~~~~~~~(f) & ~~~~~~~(g) &~~~~~~~(h)\\
\raisebox{6.5\height}{MCA-Identity}&
\epsfig{file=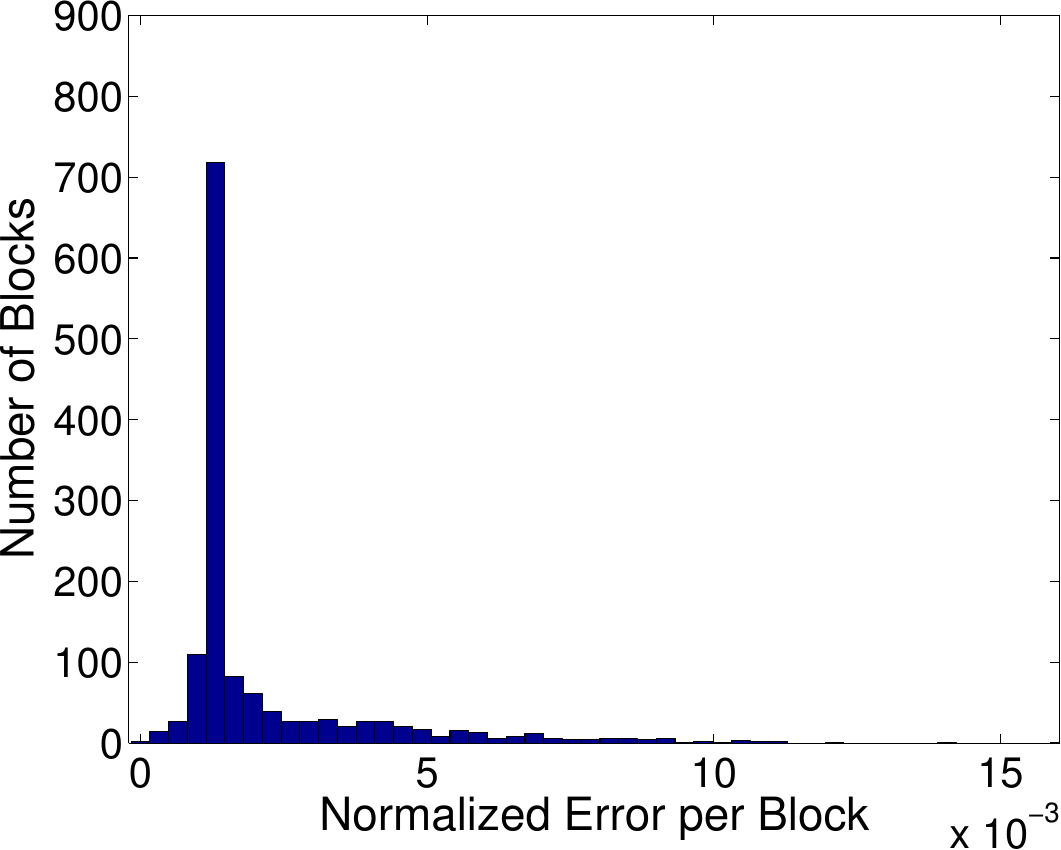,,width=0.19\linewidth,clip=} &
\epsfig{file=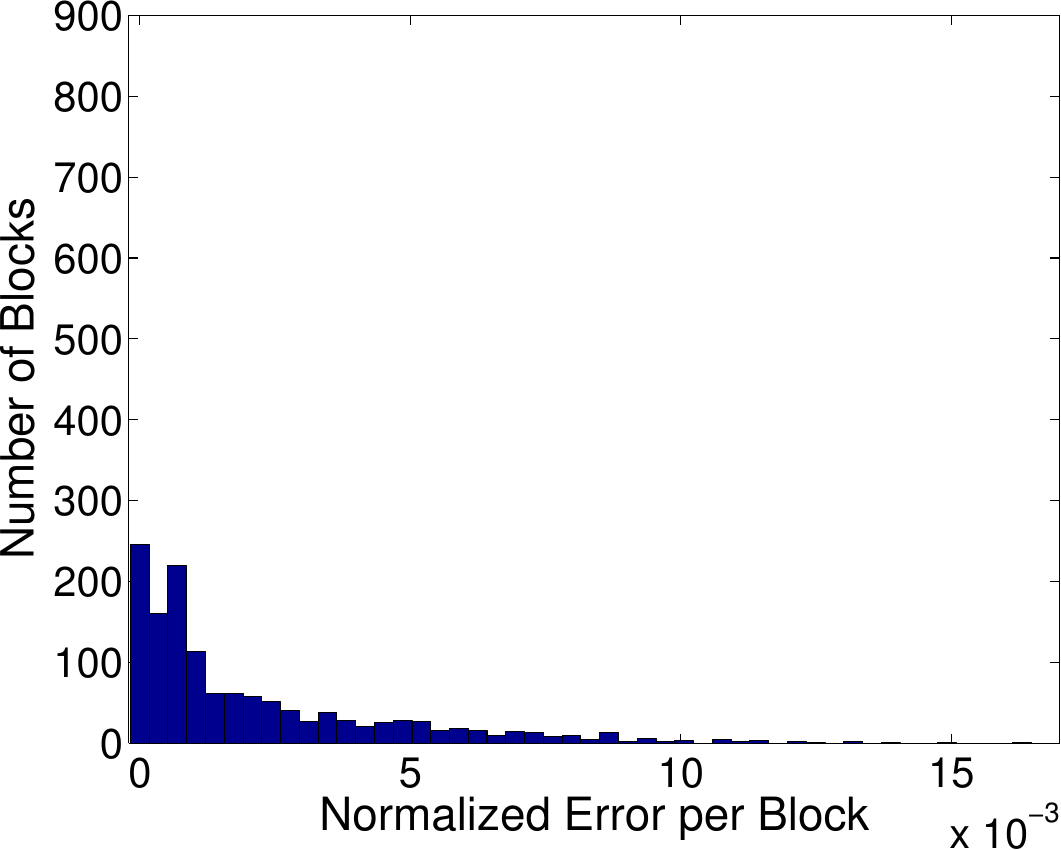,,width=0.19\linewidth,clip=} &
\epsfig{file=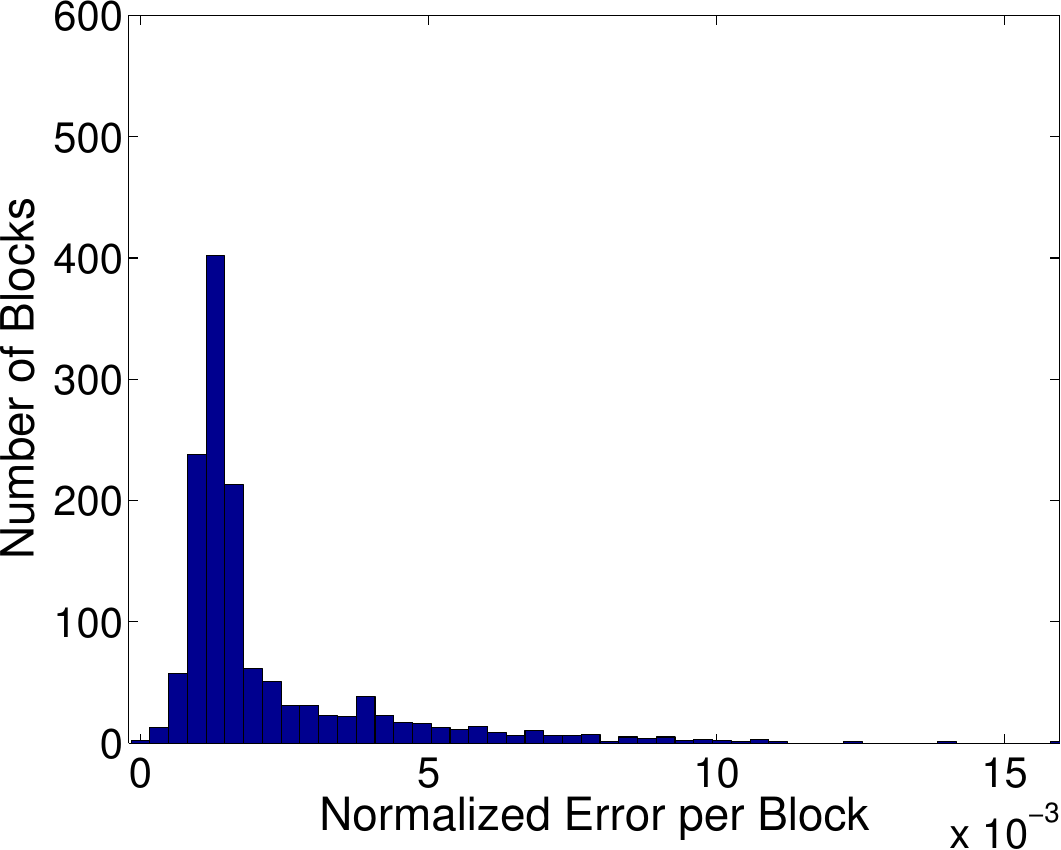,,width=0.19\linewidth,clip=} &
\epsfig{file=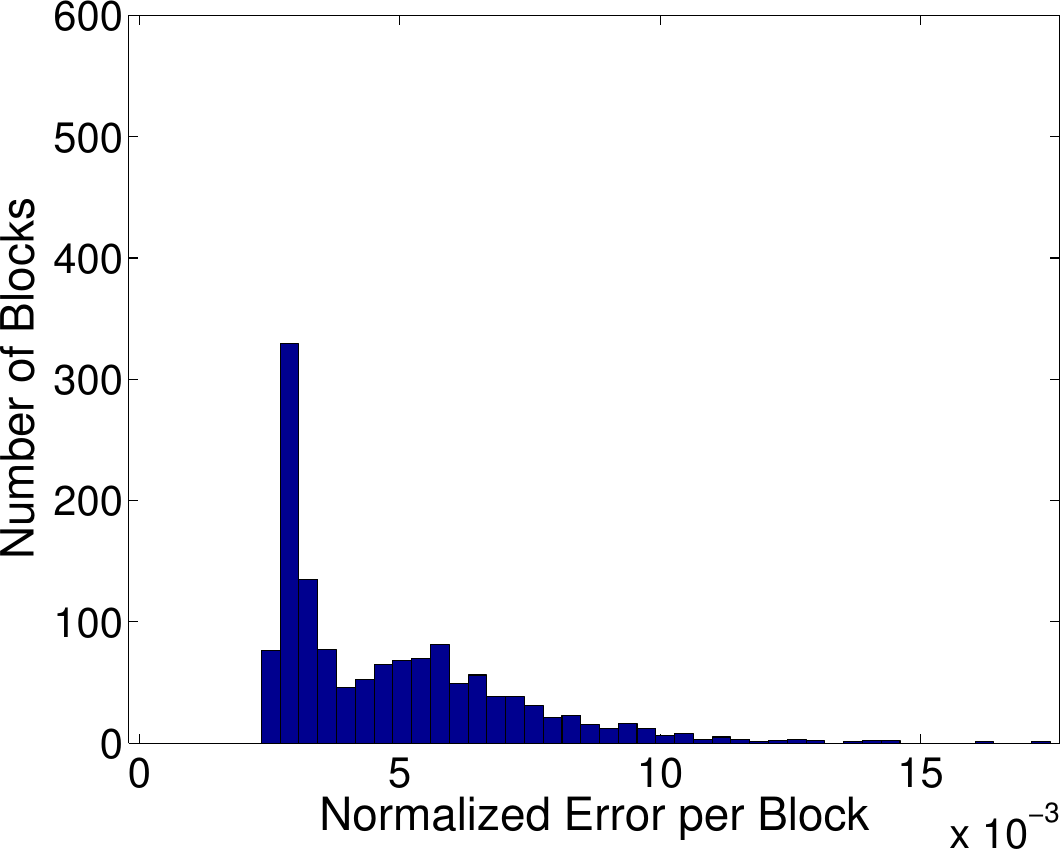,,width=0.19\linewidth,clip=}  \\
~&~~~~~~~(i) & ~~~~~~~(j) & ~~~~~~~(k) &~~~~~~~(l)\\
\raisebox{6.5\height}{NNSC (Spectral)}&
\epsfig{file=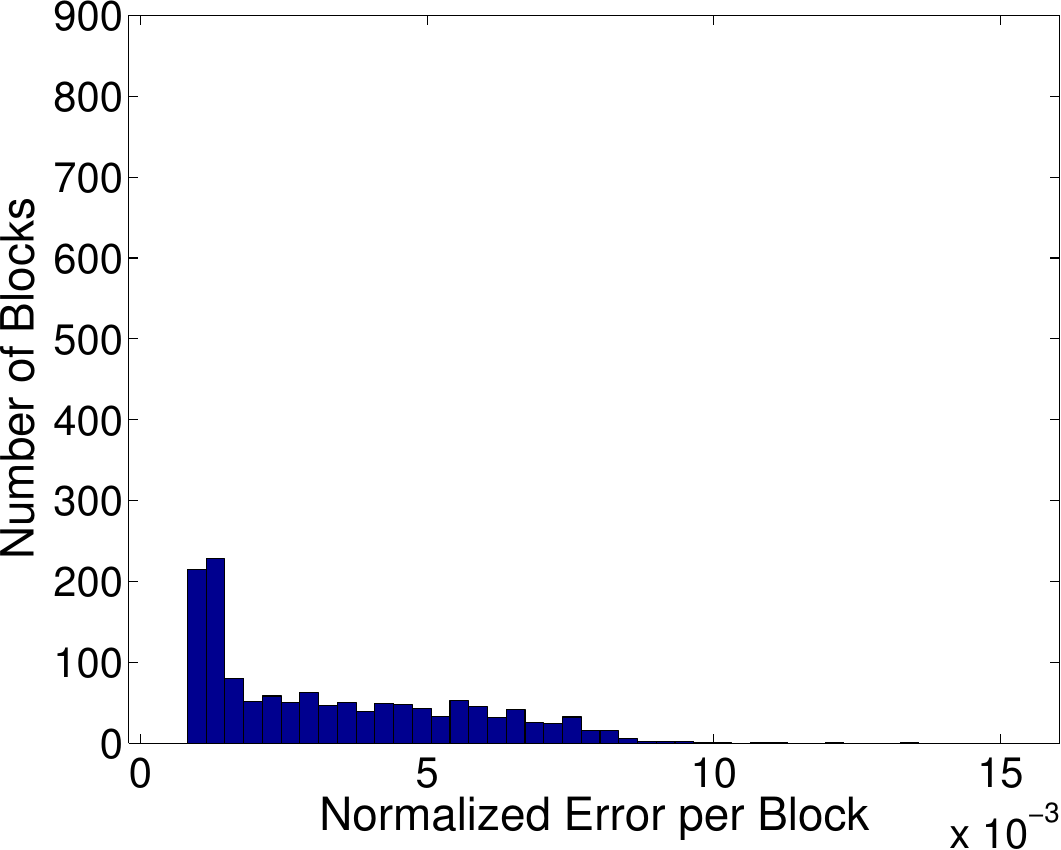,,width=0.19\linewidth,clip=} &
\epsfig{file=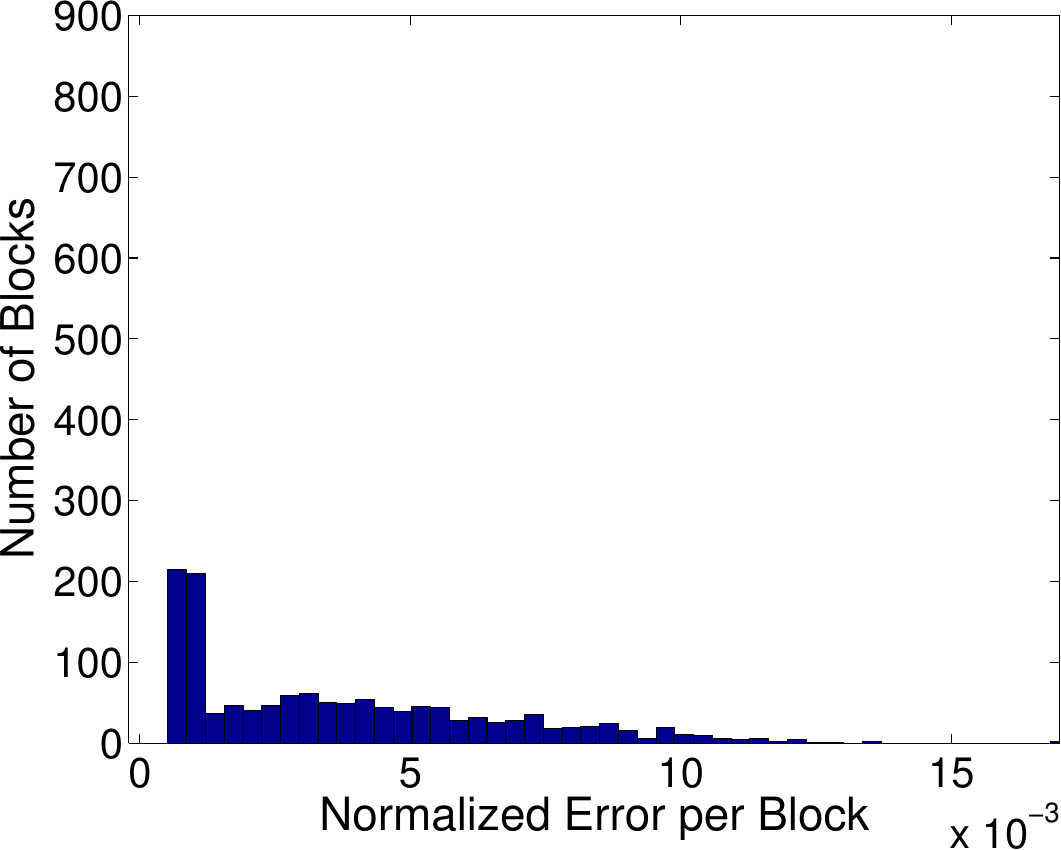,,width=0.19\linewidth,clip=} &
\epsfig{file=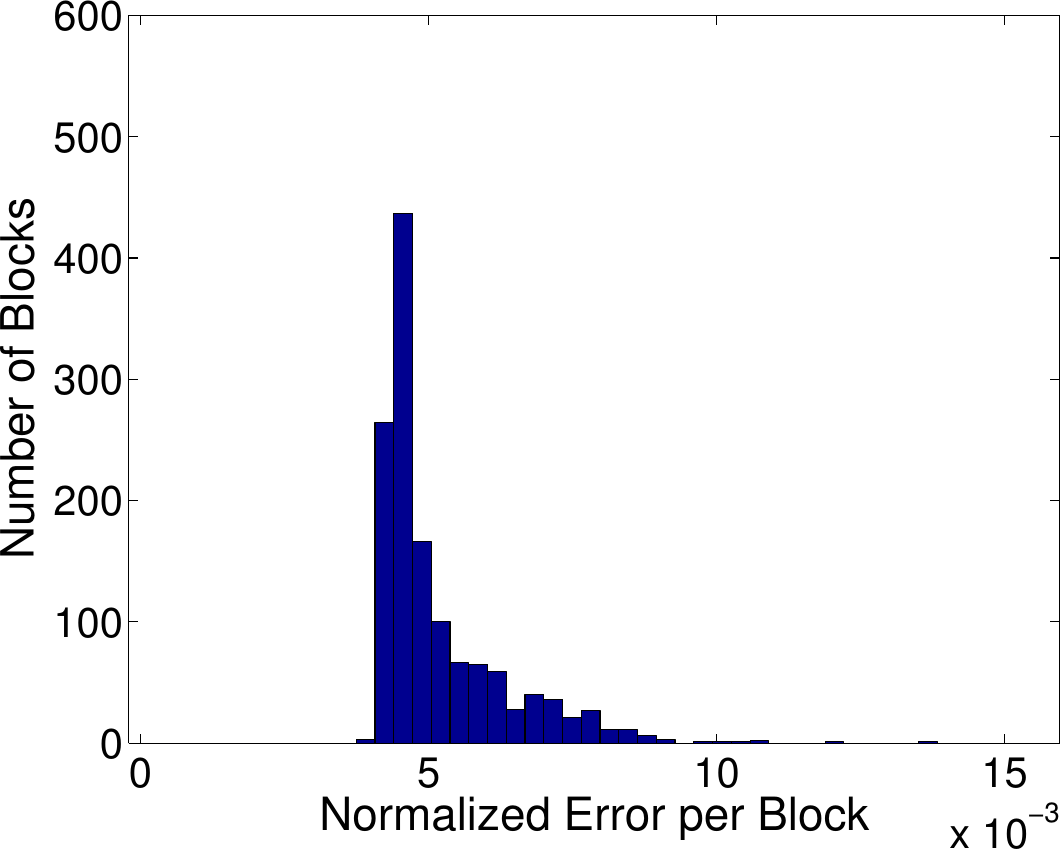,,width=0.19\linewidth,clip=} &
\epsfig{file=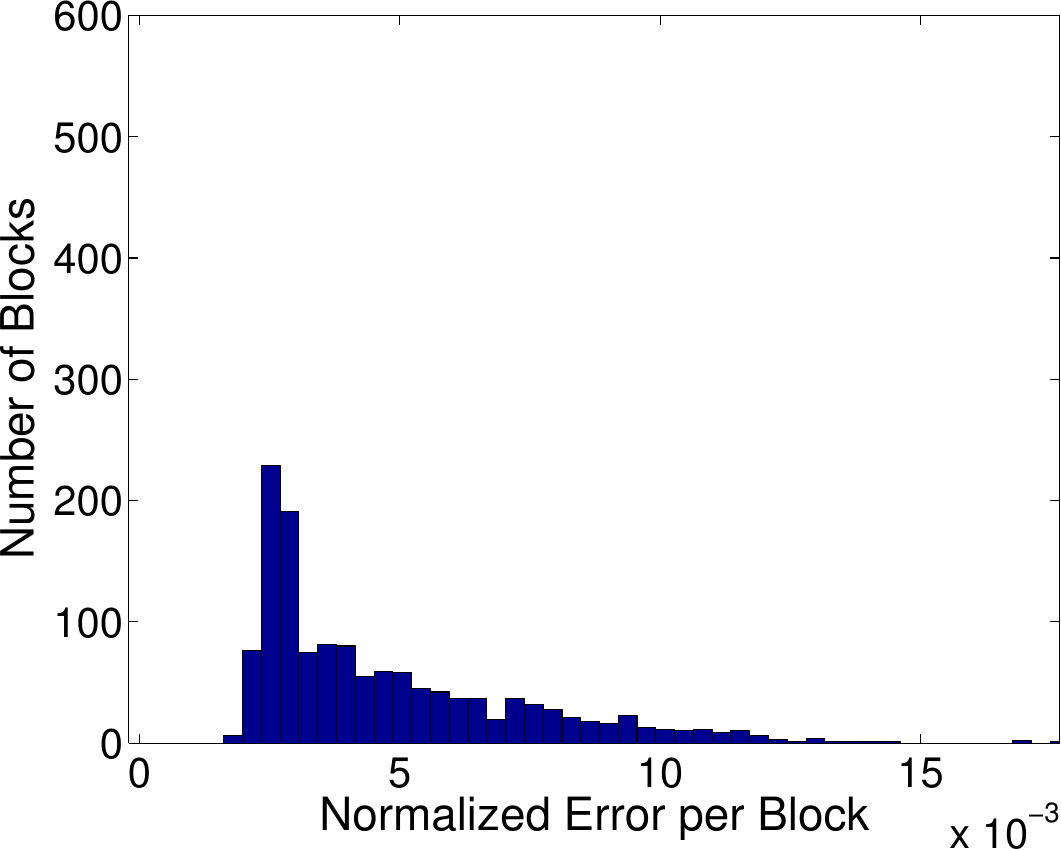,,width=0.19\linewidth,clip=}  \\
~&~~~~~~~(m) & ~~~~~~~(n) & ~~~~~~~(o) &~~~~~~~(p)\\
\end{tabularx}
\caption{ Histogram of normalized error-per-block measured using the vector $l_2$-norm of extracted nominally periodic signal $x_p$ and extracted speech signal $x_u$  for Semi-blind MCA, and MCA-DCT, and MCA-Identity, for the stylized audio forensics application. }
 \label{HistogramPlots}
\end{figure*}

A second, perhaps more illustrative, performance comparison is shown Figure~\ref{HistogramPlots}, which depicts the histogram of normalized (i.e., per-sample) errors per block, measured using the vector $\ell_2$-norm, for each method\footnote{Panels (a), (e), (i), and (m) represent the histogram of normalized error-per-block for $x_p$  and (b), (f), (j), and (n) represent the histogram of normalized error-per-block for $x_u$ via SBMCA, MCA-DCT, MCA-Identity, and NNSC (Spectral) respectively, with standard deviation of gaussian noise $\sigma=0$. Panels (c), (g), (k), and (o) represent the histogram of normalized error-per-block for $x_p$  and (d), (h), (l), and (p) represent the histogram of normalized error-per-block for $x_u$ via SBMCA, MCA-DCT and MCA-Identity respectively, with standard deviation of gaussian noise $\sigma=0.1$.}. We observe from the distribution of $\ell_2$-errors across blocks, that the SBMCA procedure (Figure~\ref{HistogramPlots} (a-d))  results in larger number of blocks with lower errors as compared to the other approaches. This feature may be of primary importance in the motivating audio forensics application where classifying each period of the nominally periodic signal $x_p$, as one of the two prototype signals, is of interest.
\vspace{-1em}

\begin{table}[t]
\caption{Comparison of reconstruction SNRs (in dB) for our proposed approach, variants of traditional MCA, and spectral separation via NMF. }
\label{SNRper}
\centering
\begin{tabular}{|l|c|c|c|c|c|}
\hline
\textbf{Noise $~\mathcal{N}(0,\sigma^2)$} & \multicolumn{2}{c|}{$\sigma=0$} & \multicolumn{2}{c|}{$\sigma=0.1$} \\
\hline

\textbf{Method $\backslash$ Signal}& $x_p$ & $x_u$ & $x_p$ & $x_u$ \\ \hline

\textbf{SBMCA }& \textbf{23.84} & \textbf{29.33}&  \textbf{20.33} & \textbf{17.14}  \\ \hline
\textbf{MCA-DCT }& 20.53 & 26.05 &  18.44 & 17.10 \\ \hline
\textbf{MCA-Identity} &11.91  & 16.25 & 11.83 & 11.96\\ \hline
\textbf{NNSC (Spectral)} & 8.89  & 13.12 & 6.57 & 11.77\\
\hline
\end{tabular}
\end{table}

\section{Discussion and Conclusions}
\label{Conclusions}

We conclude with a few additional comments related to our experimental demonstration, and more broadly, to the philosophy underlying our proposed model-based separation strategy.  In the context of audio processing, spectral source separation approaches (based on variants of NNSC and NMF) remain among the most popular techniques. Indeed, prior efforts have examined semi-blind separation techniques that are qualitatively very similar to the approach proposed here,  which aim to separate a source with known spectral content or properties from another unknown source, whose local spectral representation is learned online from the mixture data itself \cite{Smaragdis, Smaragdis2}.  Generally speaking, however, spectral-domain separation approaches find most utility in settings where the signals are, in effect, nearly orthogonal in the frequency domain, or at least in cases where the spectral overlap in minimum.  Significantly overlapping spectra in the signals being separated is widely noted as a challenging setting for these traditional methods (e.g., as noted in \cite{Smaragdis}).

On the other hand, sparse modeling and dictionary learning approaches implicitly allow for separation on the basis of overcomplete representations of each source, and ultimately of the mixture overall.  This additional modeling flexibility offers the promise that even signals with overlapping frequency domain amplitude spectra may still be separated, provided other appropriate notions of structure (sparsity in appropriate dictionaries) are employed.  Further, such modeling approaches are not restricted to the frequency domain; separation in the time domain is a viable approach if that is the domain in which the structure of the sources is most easily modeled.

Overall, encouraged by the experimental investigation here, we feel that semi-blind ``dictionary based'' separation strategies may find applications in other domains (e.g., in image or video processing) provided the structure is modeled in the appropriate domains or representations. We defer the further examination of our approach in these other application domains to future efforts.

\bibliographystyle{IEEEbib}
\bibliography{referBSS}

\end{document}